\documentclass[11pt]{iopart} 

\usepackage{hyperref}
\usepackage{graphicx}   
\usepackage{latexsym}   %
\usepackage{color}      %
\usepackage{dcolumn}    
\usepackage{bm}         
\usepackage{amssymb}    
\usepackage{graphics}
\usepackage{epsfig}
\usepackage{caption}
\usepackage{subfigure}
\usepackage{nicefrac}
\usepackage{booktabs}
\usepackage{siunitx}
\usepackage{cite}

\def\url#1{}

\begin{document}

	\title[A Data-Driven Approach]{Multi-level stochastic refinement for complex time series and fields: A Data-Driven Approach}

	\author{M. Sinhuber$^1$, J. Friedrich$^{2,3}$, R. Grauer$^2$ and M. Wilczek$^{1,*}$}
	\address{$^1$ Max Planck Institute for Dynamics and Self-Organization, Am Fa\ss berg 17, D-37077 G\"ottingen, Germany}
	\address{$^2$ Theoretische Physik I, Ruhr-Universit\"{a}t Bochum, Universit\"{a}tsstr. 150,
		D-44780 Bochum, Germany}
	\address{$^3$ Univ. Lyon, ENS de Lyon, Univ. Claude Bernard, CNRS, Laboratoire de Physique, F-69342, Lyon, France}
	\address{$^*$ Corresponding author, michael.wilczek@ds.mpg.de}

	\begin{abstract}
		Spatio-temporally extended nonlinear systems often exhibit a remarkable complexity in space and time. In many cases, extensive datasets of such systems are difficult to obtain, yet needed for a range of applications. Here, we present a method to generate synthetic time series or fields that reproduce statistical multi-scale features of complex systems. The method is based on a hierarchical refinement employing transition probability density functions (PDFs) from one scale to another. We address the case in which such PDFs can be obtained from experimental measurements or simulations and then used to generate arbitrarily large synthetic datasets. The validity of our approach is demonstrated at the example of an experimental dataset of high Reynolds number turbulence.

	\end{abstract}


	\maketitle

	\section{Introduction}\label{sec:intro}

	Many high-dimensional nonlinear systems display a remarkable degree of complexity in space and time. As a consequence, spatio-temporal records of such systems often exhibit non-trivial statistical features, including multi-scale correlations and scale-dependent deviations from Gaussianity. In principle, a comprehensive statistical characterization of such systems requires obtention of the full multi-time--multi-point statistics -- a prohibitive task given the dimensionality of typical systems.

	Significant simplifications arise if a hierarchy of scales in space and/or time can be identified, which is statistically characterized by transition probabilities from larger to smaller scales (or vice versa). If the statistics on one scale only depend on the previous one, the system is Markovian in scale. This property has been validated empirically for many complex systems, including hydrodynamic turbulence~\cite{friedrich-peinke:1997,renner01jfm,stresing10njp}, solar wind turbulence~\cite{strumik08pre}, neuronal spike trains~\cite{Sherry1982}
	and currency exchange markets~\cite{friedrich00prl,Bassler2007}. For scale-Markovian systems, the transition probabilities contain the full statistical information of the system, leading to a tremendous reduction of complexity. Even if the Markov property is weakly violated or a priori unknown, a Markovianization in scale can still provide a valuable approximation, containing, for example, non-trivial multi-scale correlations.

	In many situations, however, a purely statistical characterization is insufficient. Rather, a complete spatio-temporal record, i.e.~a realization of a complex system, is needed. For example, machine learning applications require very large training datasets. Providing such datasets from experimental measurements, computer simulations or historical records is often challenging, sometimes even impossible. Even if data is available, it sometimes lacks resolution for a specific target application due to experimental or computational constraints, such that refinement is necessary.

	In this work, we introduce a data-driven method, which allows generating statistically well-defined time series or fields with non-trivial features in a hierarchical fashion. The key idea of this \emph{multi-level stochastic refinement} (MLSR) is to generate a realization of a stochastic process representing the multi-scale features of a generic complex system given the knowledge of transition probability density functions (PDFs) on all relevant scales. Our emphasis is on a refinement of coarse time series or fields in scale, which complements previous approaches that focused on forward-in-time construction of synthetic time series by utilizing Markov properties in scale in conjunction with a Fokker-Planck-equation based approach \cite{nawroth-peinke:2006,Behnken2020}.
	We here focus on the case in which the transition PDFs can be sourced from experimental measurements or simulations.

	As one prototypical example of a complex system, we apply our approach to turbulent flows, which are ubiquitous in nature and engineering applications. Turbulent flows feature a broad range of dynamically active scales in space and time.
	For example, the separation of the large, energy-containing length scales $L$ and the small, dissipative length scales $\eta$ increases in hydrodynamic turbulence with the Reynolds number like $ L/\eta \sim Re^{3/4}$. As a consequence, fully resolved simulations are computationally prohibitive for realistic natural and engineering flows. Large-eddy simulations tackle this problem by only resolving the energy-containing scales~\cite{meneveau2000arfm,sagaut06}. However, in many applications, resolving small-scale turbulence is key.
	Examples include cloud microphysics~\cite{shaw2003,grabowski2013growth,pumir2016collisional}, turbulent combustion~\cite{Meneveau1991b,Dopazo2015},
	mixing and transport of atmospheric pollutants~\cite{Gifford1982,vallis2017atmospheric}, the assessment of turbulent loads on wind turbines \cite{Muecke2011}, as well as the development of active control strategies for wind farms \cite{Spencer2013}. Recently, deep learning algorithms for feature identification and extraction have been gaining attention to deal with these challenges. For the required large amounts of training data~\cite{Foresti2011,Heinermann2016}, synthetic turbulent fields, like the one presented in the following, could make a valuable contribution to alleviate the challenge of acquiring the databases.

	The paper is structured as follows: We start with a description of the algorithm and the theoretical background in section~\ref{sec:threepointpdf}.
	We then demonstrate the data-driven algorithm at the example of state-of-the-art wind tunnel measurements of high-Reynolds number turbulence in section~\ref{sec:data-driven-turbulence}. In the conclusions~\ref{sec:conclusions} we discuss future applications and generalizations as well as an outlook on model-driven approaches to the MLSR method.

	\section{Description of the data-driven MLSR based on three-point PDFs}
	\label{sec:threepointpdf}

	\begin{figure}
		\begin{center}
			\includegraphics[width=0.8\textwidth]{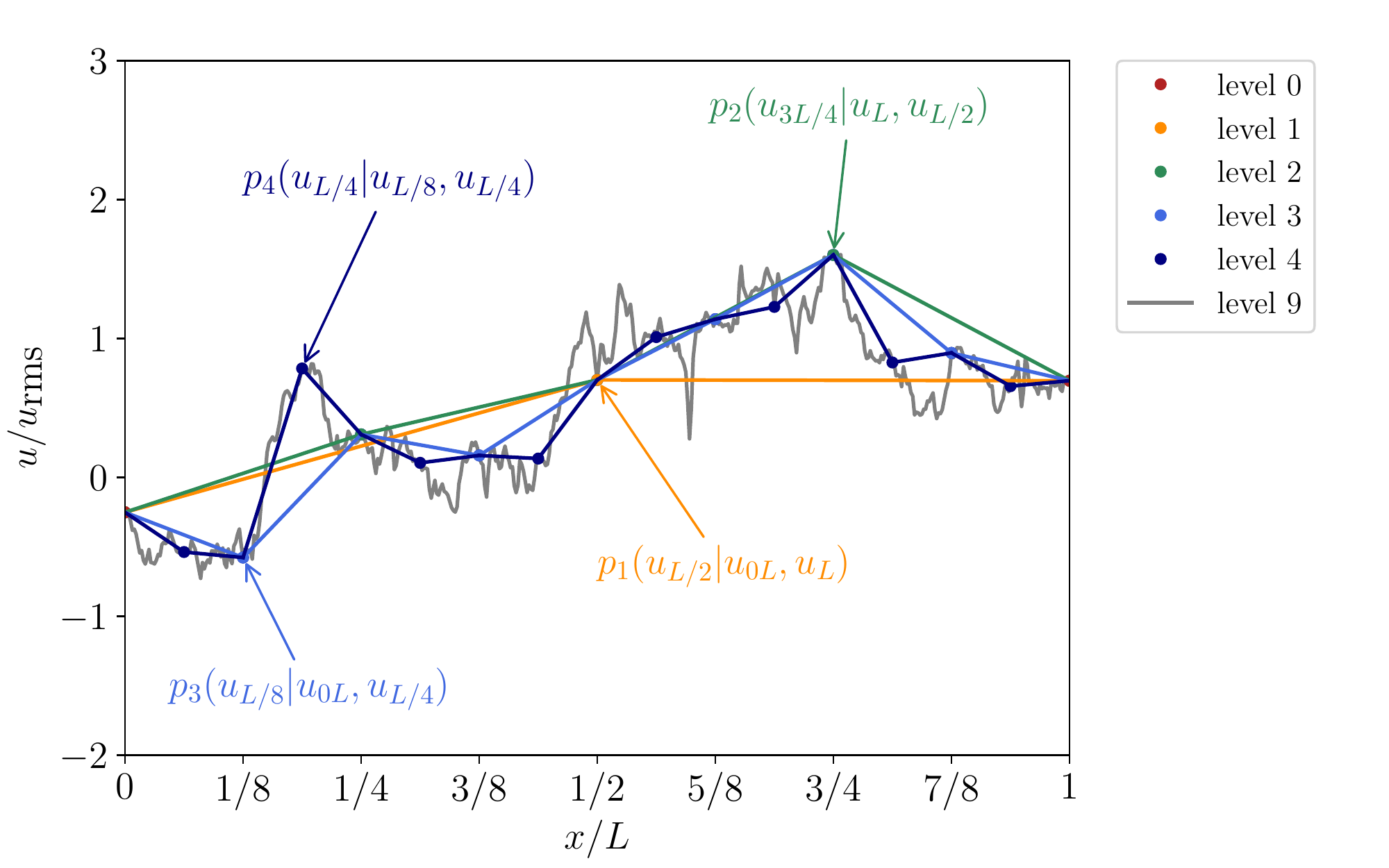}
		\end{center}
		\caption{Schematic representation of the data-driven multi-level stochastic refinement at the example of synthetic time series (mapped to a spatial signal using Taylor's hypothesis \cite{Taylor1938}) of wind tunnel turbulence. The initial boundary values are set on a scale of $L=32\lambda$ ($\lambda$ denotes the Taylor scale in hydrodynamic turbulence). The velocity at $L = 16\lambda$, i.e.~on half the scale, is then drawn from the the conditional PDF $p(u_{L/2} | u_{0L}, u_{L} )$ which previously has been obtained from experimental data. By iterating this procedure, the synthetic time series (gray) is obtained, which shares statistical features with the original experimental data.}
		\label{fig:mlsr_schematic}
	\end{figure}

	Let us consider the situation in which we aim to generate a realization of a statistically homogeneous, one-dimensional field of length $L$ with a resolution of $\Delta = L/2^N$. Given initial field values on scale $r_0=L$, we adopt a hierarchical refinement on scale $r_1=L/2$, then progressing to scale $r_2=L/4$ etc.~until scale $r_N=\Delta=L/2^N$ is reached after $N$ refinement steps.

	On each refinement level, let us assume we know the three-point PDF $f(u_{\mathrm l},u_{\mathrm c},u_{\mathrm r};r_i)$ from experimental measurements. Here $u_{\mathrm l}$, $u_{\mathrm c}$ and $u_{\mathrm r}$ denote the field values at the left end, in the center, and at the right end of the interval, respectively, which are separated by the distance $r_i$. Based on the three-point PDF, we can define the conditional PDF
	\begin{equation}\label{eq:conditionalpdf}
	p_i\left(u_{\mathrm{c}}|u_{\mathrm{l}}, u_{\mathrm{r}}\right) \equiv p\left(u_{\mathrm{c}}|u_{\mathrm{l}}, u_{\mathrm{r}}, r=r_i\right) = \frac{f(u_{\mathrm l},u_{\mathrm c},u_{\mathrm r};r_i)}{f(u_{\mathrm l},u_{\mathrm r};2r_i)} \, .
	\end{equation}
	Here, $f(u_{\mathrm l},u_{\mathrm r};2r_i)$ denotes the two-point PDF on scale $2r_i$, which can be obtained from the three-point PDF by integrating out the center point:
	\begin{equation}
	f(u_{\mathrm l},u_{\mathrm r};2r_i) = \int \mathrm{d}u_{\mathrm c} \, f(u_{\mathrm l},u_{\mathrm c},u_{\mathrm r};r_i) \, .
	\end{equation}
	The conditional PDF \eref{eq:conditionalpdf} can be interpreted as a transition PDF in scale. It describes the probability of drawing $u_{\mathrm{c}}$, which can be used to obtain a field which is sampled on scale $r_i$, given the two values $u_{\mathrm{l}}$ and $u_{\mathrm{r}}$ at the end points of the interval of length $2r_i$. In our numerical implementation, we employ a standard rejection method \cite{vonNeumann1951} to draw from this conditional PDF. We thereby construct a realization sampled on the scale $r_i$. The three-point statistics on scale $r_i$ are consistent with the three-point statistics of the underlying experimental data by construction. Employing this multi-level stochastic refinement algorithm hierarchically starting from $r_1$ and progressing down to scale $r_N$ enforces consistency with the underlying data in each refinement step.

We illustrate the procedure in figure \ref{fig:mlsr_schematic} for a time series of flow velocities in turbulence. Here we construct a synthetic turbulent velocity signal at a resolution of $L/2^9=\lambda/16$ over an interval of length $L=32\lambda$, where $\lambda$ denotes the Taylor scale of hydrodynamic turbulence. As detailed in section \ref{sec:data-driven-turbulence}, the corresponding transition PDFs have been obtained from wind tunnel measurements reaching down to scales $L/2^9=\lambda/16$. Starting from prescribed velocity values at the end points of the interval $[0,L]$, the velocity $u_{L/2}$ at the mid-point is drawn from the transition PDF $p_1(u_{\mathrm{c}} | u_{\mathrm{l}}=u_{0L}, u_{\mathrm{r}}=u_{L})$. The two velocities $u_{L/4}$ and $u_{3L/4}$ on the next level of refinement are drawn from $p_2(u_{\mathrm{c}} | u_{\mathrm{l}}=u_{0L}, u_{\mathrm{r}}=u_{L/2})$ and $p_2(u_{\mathrm{c}} | u_{\mathrm{l}}=u_{L/2}, u_{\mathrm{r}}=u_{L})$, respectively. The four velocities necessary to refine the field on scale $r_3=L/8$ are then drawn from the corresponding transition PDF $p_3$. By iterating this procedure down to level $9$, we finally obtain the gray time series shown in figure \ref{fig:mlsr_schematic}.

	Up to now, we have discussed how to generate a field of length $L$ and resolution $L/2^N$. If the scale $L$ is large enough such that field values separated by $L$ are statistically independent, arbitrarily large fields can be generated by drawing from the single-point PDF $f(u)$ on scale $L$. Statistical correlations on scales smaller than $L$ are then subsequently introduced by MLSR. Regarding the small-scale resolution, the latter is only limited by the smallest experimentally accessible scale, from which the transition PDFs are estimated.

	\subsection{Multi-point single-time statistics and relation to Markov processes in scale}
	\label{sec:relationtomarkov}

	The MLSR discussed in the previous section allows us to generate a synthetic field of extent $L$ and resolution $\Delta = L/2^N$ sampled on $M+1$ points where $M=2^N$. Let us denote the field values at these points sequentially by $u_0 \dots u_M$ with the corresponding multi-point PDF $f(u_0,u_1,u_2,u_3,\dots,u_M)$. Owing to the structure of the MLSR algorithm, an ensemble of such synthetically generated fields obeys the multi-point statistics
	\begin{equation}\label{eq:jointpdf}
	\fl  f(u_0,u_1,u_2,u_3,\dots,u_M) = f(u_0,u_M) \, \prod_{i=1}^N \prod_{j=1}^i p_i\big( u_{(2j-1)M/2^i} \big| u_{(2j-2)M/2^i}, u_{2jM/2^i} \big) \, .
	\end{equation}
	Consistent with the hierarchical construction of the random fields, this expression for the multi-point statistics shows explicitly that random variables can be systematically integrated out level by level. For example, by integrating out all random variables which have been introduced in the $N$th refinement step resulting in a field resolution of $\Delta$, the joint PDF $f(u_0,u_2,u_4,\dots,u_M)$ corresponding to an ensemble of fields with a resolution of $2\Delta$ is obtained.

The hierarchical structure of the multi-point statistics \eref{eq:jointpdf}, in which the probability of the field value at the center point of an interval is determined by the values at the two end points, indicates that MLSR generates a Markov process in scale.
As we show by direct calculation in \ref{sec:markovappendix}, Markovian properties also follow for related statistical quantities such as field increments.
As a further note, if the original time series obeys Markov properties of the form $p\left(u_1|u_0,u_2,u_3\right)=p\left(u_1|u_0,u_2\right)$, then equation \eref{eq:jointpdf} is an exact representation of the multi-point statistics. For turbulent flows, this property is an ongoing topic of investigation \cite{Peinke2020}.

	A prototypical example for non-Gaussian, yet Markovian, systems are time series generated by nonlinear Langevin equations. For such time series, the MLSR algorithm is capable of generating synthetic equivalents that reproduce the correct $N$-point statistics precisely. As a proof on concept, we demonstrate this for the two-point statistics in \ref{sec:data-driven-langevin}. However, even if Markovian properties are mildly violated, the synthetic time series generated by MLSR still resemble the original statistics closely as illustrated in section \ref{sec:data-driven-turbulence}.

	\section{Application: Turbulence Data}\label{sec:data-driven-turbulence}

	Previous works (e.g.~\cite{friedrich-peinke:1997,renner01jfm,stresing10njp}) have empirically established a Markov property in scale for hydrodynamic turbulence, which lends support to the MLSR approach proposed here to capture essential statistical features of turbulent fields. To illustrate the capabilities of the MLSR method for real-world, close-to-Markovian complex systems, we utilize experimental turbulence data from the Variable Density Turbulence Tunnel VDTT at the Max Planck Institute for Dynamics and Self-Organization \cite{Bodenschatz2014}. This pressurizable wind tunnel uses sulfur hexafluoride at densities up to 15bar to achieve high Reynolds numbers due to the low kinematic viscosity of the pressurized working gas. The data we used to construct the MLSR method is a subset of the data presented in \cite{Sinhuber2017}. Turbulence was generated using a passive grid of rectangular grid bars with a mesh spacing of 18cm and was measured about 40 mesh sizes downstream of the grid using a nano-scale thermal anemometry probe (NSTAP) developed at Princeton University \cite{Bailey2010, Vallikivi2011}. The NSTAP was a 60$\mu$m long hot-wire that measured one-dimensional velocity signals and was small enough to resolve the smaller length scales in the VDTT. The data was obtained at a pressure of 8.5bar of sulfur hexafluoride, resulting in a Taylor-scale Reynolds number of $R_\lambda=1030$. The time series was sampled at a rate of 60kHz with noise being removed at 15kHz using an 8\textsuperscript{th}-order Butterworth filter and contained $\mathcal{O}\left(10^{10}\right)$ samples of the turbulent velocity, corresponding to $\mathcal{O}\left(10^5\right)$ large eddy turnover times. Based on the mean speed $U=4.2$m/s and the root mean square velocity fluctuation $u_\mathrm{rms}=0.13$m/s, the turbulence intensity $u_\mathrm{rms}/U$ was 3\%. We invoked Taylor's hypothesis \cite{Taylor1938} to convert temporal information into spatial information. The Kolmogorov length scale, $\eta$, was 34$\mu$m, the integral length scale, $L_{\mathrm{int}}$, was 12.6cm and the Taylor scale, $\lambda$, was 2.2mm.

To obtain the conditional PDFs $p\left(u_{\mathrm{c}}|u_{\mathrm{l}}, u_{\mathrm{r}}, r \right)$, we first subtracted the mean of the signal velocity and computed the three-point PDF $f(u_{\mathrm l},u_{\mathrm{c}},u_{\mathrm{r}};r)$ and the two-point PDF $f(u_{\mathrm l},u_{\mathrm{r}};2r)$, using conventional binning with 100 equidistant bins in all PDF dimensions. Given the substantial length of the velocity time series, these multi-point PDFs are well-converged. In situations where one potentially only has access to shorter time series, the estimates of the PDFs could be improved using e.g.~kernel density estimation
	\cite{Gramacki2018} at a higher computational cost. To remove spurious noise stemming from divisions of bin elements with insufficient statistics when computing conditional statistics, we empirically set all bins in the three-point PDF that contain less than 10 samples of the turbulent velocity to zero.

	\begin{figure}
		\centering
		\includegraphics[]{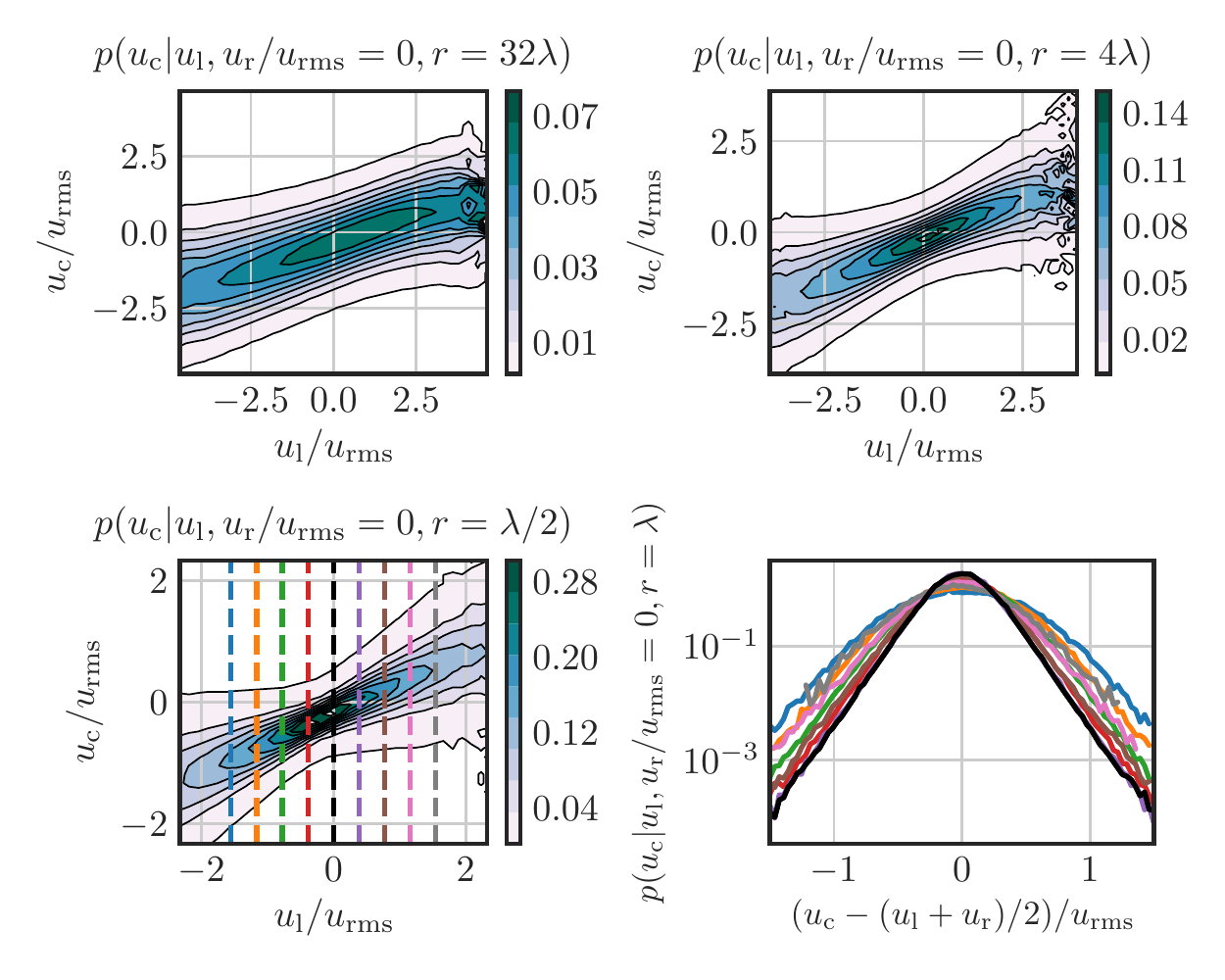}
		\caption{Cuts through the conditional PDF $p(u_{\mathrm{c}}|u_{\mathrm{l}},u_{\mathrm {r}},r)$ for $u_\mathrm{r}/u_\mathrm{rms}=0$ and for separations $r$ of 0.5$\lambda$, 4$\lambda$, 32$\lambda$. With decreasing separation, the shape of conditional PDFs qualitatively changes, showing a more dominant constriction for $u_\mathrm{l}= u_\mathrm{r}$. Bottom right: One-dimensional cuts through the conditional PDFs for $u_\mathrm{r}/u_\mathrm{rms}=0$ and fixed $u_\mathrm{l}$ indicated by the line color. For large differences $\left|u_\mathrm{l}-u_\mathrm{r}\right|$, the PDFs are almost Gaussian, while significant non-Gaussianity emerges with decreasing difference $\left|u_\mathrm{r}-u_\mathrm{l}\right|$. Irrespective of the difference $\left|u_\mathrm{l}-u_\mathrm{r}\right|$, all PDFs are centered around the mean $\left(u_\mathrm{l}+u_\mathrm{r}\right)/2$.}
		\label{fig:condpdf}
	\end{figure}

	Two-dimensional cuts through these conditional PDFs computed from the NSTAP data for $u_{\mathrm{r}}/u_\mathrm{rms}=0$ are shown in figure \ref{fig:condpdf}. At small scales $r$, the conditional PDF is noticeably constricted around $u_\mathrm{l}/u_\mathrm{rms} = u_\mathrm{r}/u_\mathrm{rms} = 0$ and substantially widens with increasing difference $\left|u_\mathrm{l}-u_\mathrm{r}\right|$. As to be expected, the constriction around $u_\mathrm{l}/u_\mathrm{rms} = u_\mathrm{r}/u_\mathrm{rms} = 0$ weakens when the separation becomes comparable to the integral length scale such that the shape of the conditional PDFs, which are centered around $\left(u_\mathrm{l}+u_\mathrm{r}\right)/2$, varies less with the difference $\left|u_\mathrm{l}-u_\mathrm{r}\right|$. One-dimensional cuts through the conditional PDFs in figure \ref{fig:condpdf} at a scale of $r = \lambda$ are shown in the bottom right panel. For small differences $\left|u_\mathrm{l}-u_\mathrm{r}\right|$ the cuts through the conditional PDFs show a noticeably non-Gaussian behavior, while for large differences $\left|u_\mathrm{l}-u_\mathrm{r}\right|$, the PDFs become virtually Gaussian. This emphasizes that the conditional PDFs not only widen with increasing differences $\left|u_\mathrm{l}-u_\mathrm{r}\right|$, but rather change shape and therefore result in a qualitative change of the underlying dynamics determining $u_\mathrm{c}$.

	\begin{figure}
		\centering
		\includegraphics[]{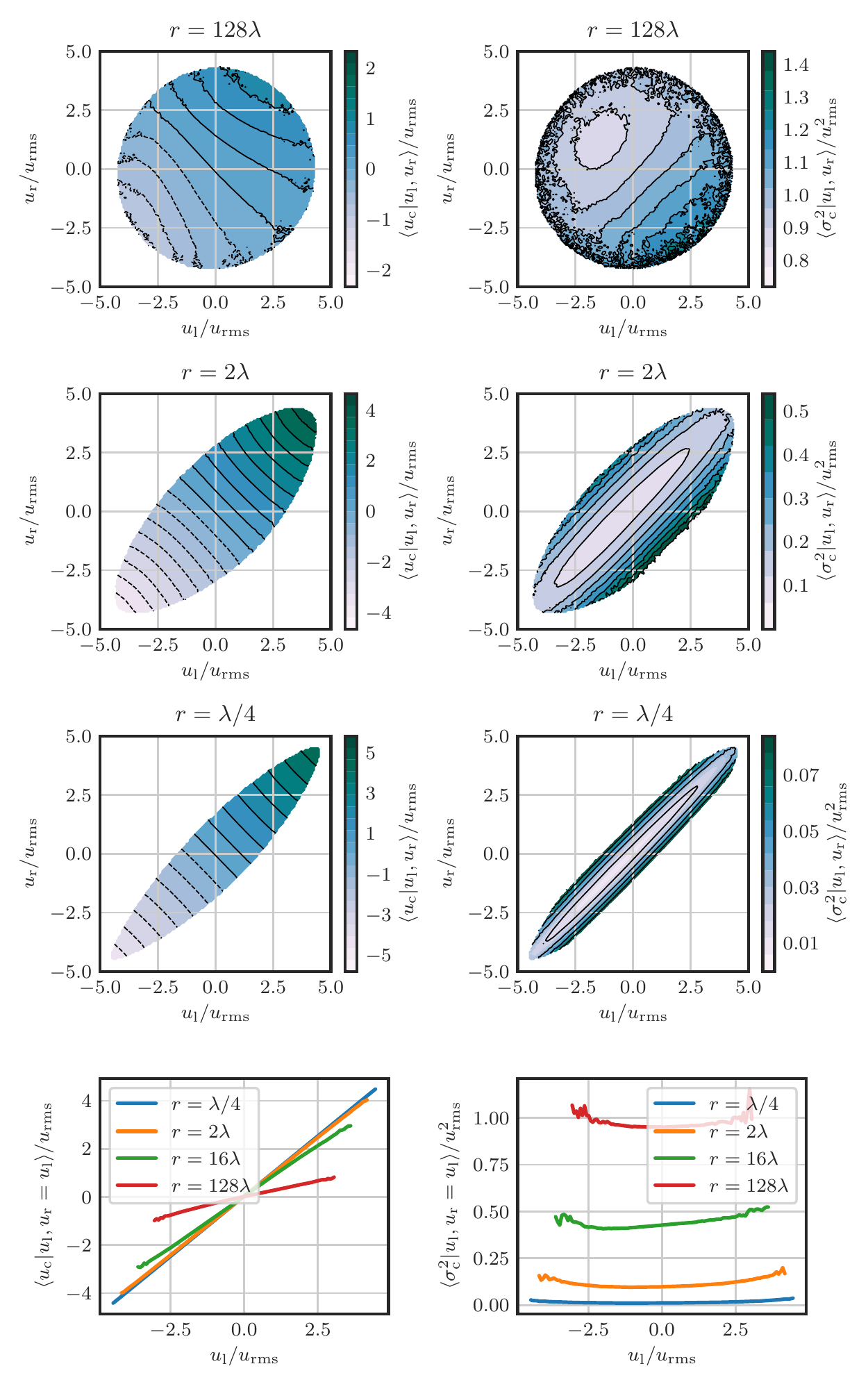}
		\caption{Left column (top to bottom): Mean of the conditional PDF $p\left(u_{\mathrm{c}}|u_{\mathrm{l}}, u_{\mathrm{r}}, r \right)$ for separations $\lambda/4$, 2$\lambda$ and 128$\lambda$ as well as one-dimensional diagonal cuts through the conditional mean for $u_{\mathrm{l}}=u_{\mathrm{r}}$. The conditional mean appears to be a linear function of the mean value $\left(u_{\mathrm{l}}+u_{\mathrm{r}}\right)/2$, with a slope that approaches one for decreasing separation. Right column (top to bottom): Conditional variance as well as one-dimensional diagonal cuts for $u_{\mathrm{l}}=u_{\mathrm{r}}$. In contrast to the conditional mean, the variance is a slightly asymmetric, non-trivial function of $u_{\mathrm{l}}$ and $u_{\mathrm{r}}$.}
		\label{fig:firstmoment}
	\end{figure}

	This effect can also be seen in the mean and variance of the conditional PDF in figure \ref{fig:firstmoment}. While the conditional mean shows a nearly linear behavior across all separations with a slope that depends on the scale $r$, the variance is non-trivially depends on the separation and the difference $\left|u_\mathrm{l}-u_\mathrm{r}\right|$. However, the deviations from symmetric behavior are relatively small.

\begin{figure}
		\centering
		\includegraphics[width=0.8\textwidth]{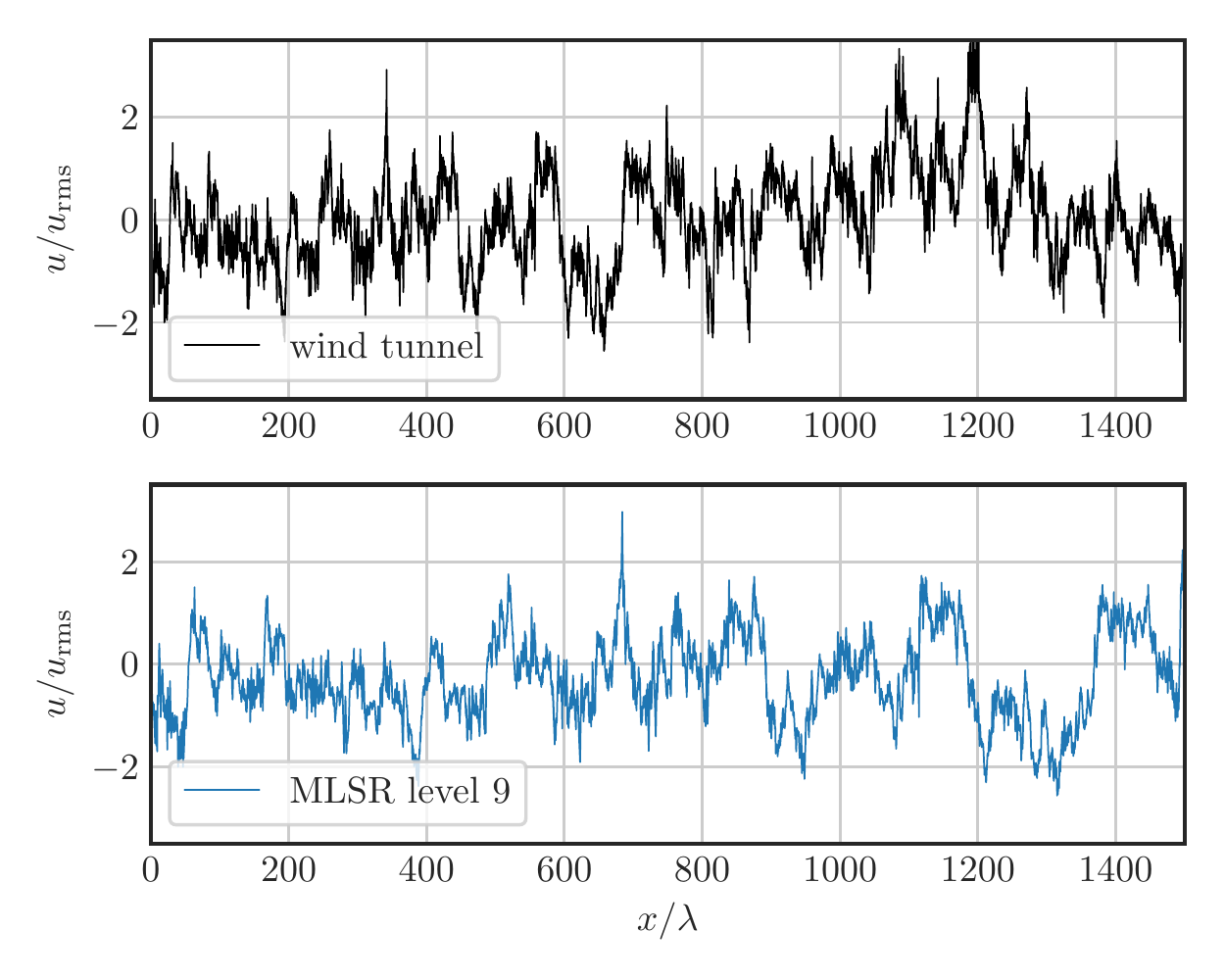}
		\caption{Comparison of time series from wind tunnel data and one obtained from a 9-level MLSR procedure.}
		\label{fig:refinement}
	\end{figure}

	Using these PDFs and the method described in section \ref{sec:threepointpdf}, we can construct a multi-level refined time series reproducing the key properties of a real turbulent time series. To initialize the method, we started with a turbulent time series sampled at $r_0 =32\lambda$ and utilized a 9-level MLSR process to reach a final separation $r_{9} = \lambda/16$. Figure \ref{fig:refinement} shows a direct comparison of the resulting synthetic time series with an experimentally obtained turbulent time series sampled at the same separation $r_{9} = \lambda/16$. Both time series show a visually indistinguishable behavior, the intermittent character of turbulence is qualitatively well reproduced by the MLSR time series.

\begin{figure}
		\centering
		\includegraphics{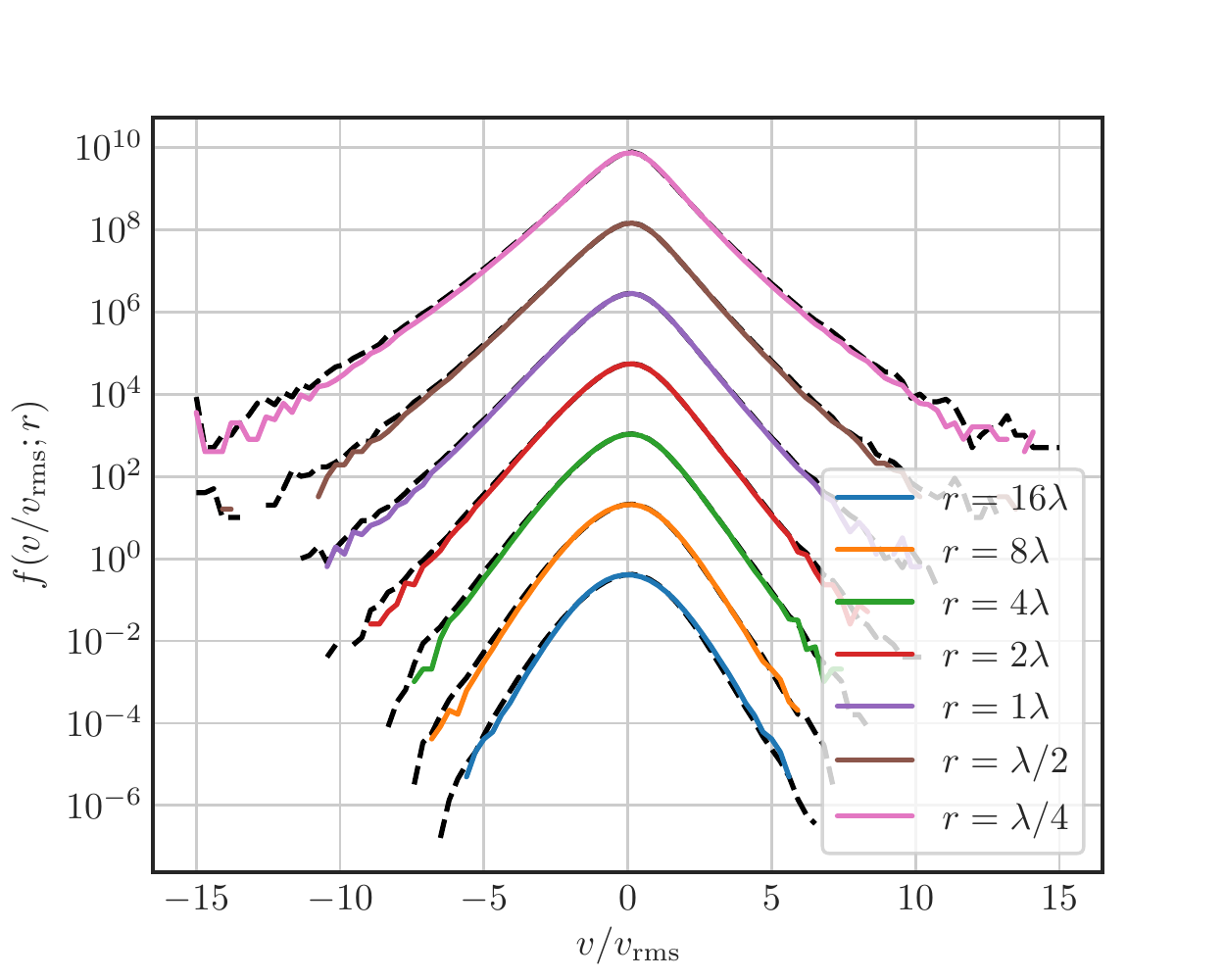}
		\caption{PDFs of the velocity increments at various separations for each refinement scale. The curves are staggered for better visibility. The black line corresponds to PDFs from the original turbulence dataset, the colored lines to PDFs obtained from the MLSR fields.}
		\label{fig:incrementpdfs}
	\end{figure}

	As a quantitative benchmark for the MLSR algorithm, we computed the velocity increment PDFs $f\left(v;r\right)$, where $v = u\left(x+r\right)-u\left(x\right)$, at each refinement scale for the respective separation as well as the velocity increment PDFs for the experimental turbulence dataset that was originally used to compute the conditional PDFs. The result is shown in figure \ref{fig:incrementpdfs}. At each refinement scale, the refined time series reproduces the original turbulence PDFs including the increasingly heavy tails towards smaller scales. As detailed in \cite{Luck2006}, the Markovian properties of turbulence start to deteriorate at scales smaller than $\lambda$, the Taylor scale. While deviations are small, effects of this violation can supposedly be seen in the tails of the increment PDFs for smaller scales in figure  \ref{fig:incrementpdfs}. This is in contrast to the findings of the exact reproduction for fully Markovian systems in \ref{sec:data-driven-langevin}. It is, however, remarkable how closely the MLSR-generated time series quantitatively resemble actual fluid turbulence highlighting its general applicability for systems without ideal Markov properties.

	\section{Conclusions and Outlook}
	\label{sec:conclusions}
	We have presented a hierarchical stochastic refinement method that allows to generate synthetic fields and time series. In this data-driven MLSR, the synthetic fields (time series) are reconstructed from multi-point (multi-time) statistics of \emph{real input data}. The method is able to generate synthetic data which exhibits two-point statistics in good agreement with the original data.	We demonstrated the data-driven MLSR at the example of turbulent wind tunnel data as well as for nonlinear Langevin equations.

	Due to the low requirements of the method, i.e.~the determination of the three-point PDF for different refinement levels, as well as due to the fast production of surrogate data, the data-driven MLSR appears to be a promising method for a broad range of applications. Synthetic fields generated by MLSR, for example, might be used as input to simulate power output fluctuations of wind parks, the fatigue loads on single wind turbines~\cite{Lind2014}, to test the stability of power systems~\cite{Schmietendorf2014}, as well as to simulate the propagation of cosmic rays \cite{schlickeiser:2015,zweibel:2013,reichherzer-tjus-zweibel-etal:2020}.

	In certain situations, sufficient statistics to resolve the three-point PDFs required for the data-driven MLSR method might not be available and alternative approaches are necessary. An extension of the approach to the generation of synthetic data based on phenomenological models of turbulence~\cite{atmos11091003,friedrich2020stochastic} without the need for measuring three-point PDFs is subject of ongoing work.

	\section*{Acknowledgments}
	We are grateful to Eberhard Bodenschatz for the opportunity of using the data acquired during the doctoral studies of MS at the Max Planck Institute for Dynamics and Self-Organization for this project as well as to Gregory P. Bewley for participating in the data acquisition and for fruitful discussions. We thank Joachim Peinke for valuable discussions.

	MW is supported by the Max Planck Society. JF acknowledges funding from the Humboldt Foundation within a Feodor-Lynen fellowship and also benefitted from the financial support of the Project IDEXLYON of the University of Lyon in the framework of the French program ``Programme Investissements d'Avenir'' (ANR-16-IDEX-0005). MS acknowledges support from the Deutsche Forschungsgemeinschaft under grant no. 396632606.

	\appendix

	\section{Relation of MLSR to Markov processes in scale}
	\label{sec:markovappendix}
	To elucidate the relation of the MLSR approach to Markov processes in scale in terms of increments, let us consider one particular ``branch" of random variables in \eref{eq:jointpdf} which connects the largest scales to the smallest scales. Without loss of generality, this can be achieved by setting $j=1$ in \eref{eq:jointpdf} which results in the reduced multi-point PDF
	\begin{equation}\label{eq:reducedjointpdf}
	\fl  f(u_0,u_1,u_2,u_4,u_8,u_{16},\dots,u_M) = f(u_0,u_M) \, \prod_{i=1}^N  p_i\big( u_{M/2^i} \big| u_0, u_{M/2^{(i-1)}} \big) \, .
	\end{equation}
	The joint PDF of increments $v_i = u_{M/2^i} - u_0$ on scales $r_i$ together with $u_0$ are obtained from the reduced multi-point PDF \eref{eq:reducedjointpdf} according to
	\begin{eqnarray}\label{eq:incrementpdf}
	\fl  &f(v_0,v_1,v_2,v_3,\dots,v_N,u_0) \\
	= &\prod_{i=0}^N \int \mathrm{d}u_{M/2^i} \, \delta\left(u_{M/2^i}-u_0-v_i\right)   f(u_0,u_1,u_2,u_4,u_8,u_{16},\dots,u_M) \, . \nonumber
	\end{eqnarray}
	After inserting \eref{eq:reducedjointpdf} into \eref{eq:incrementpdf} the integration can be carried out and we obtain
	\begin{equation}
	\fl  f(v_0,v_1,v_2,v_3,\dots,v_N,u_0) =  f(u_0,v_0+u_0) \, \prod_{i=1}^N  p_i\big( v_i+u_0 \big| u_0, v_{i-1}+u_0 \big) \, .
	\end{equation}
	Similar to the multi-point PDFs, the increment variables can be integrated out scale by scale. For example, the joint PDF which includes all increments but the smallest,
	\begin{equation}\label{eq:nincpdf}
	\fl  f(v_0,v_1,v_2,v_3,\dots,v_{N-1},u_0) =  f(u_0,v_0+u_0) \, \prod_{i=1}^{N-1}  p_i\big( v_i+u_0 \big| u_0, v_{i-1}+u_0 \big) \, ,
	\end{equation}
	can be straightforwardly obtained by integrating over $v_N$. As a consequence, the conditional increment PDF fulfills a Markov property, which can be seen by noticing that \eref{eq:nincpdf} and \eref{eq:n-1incpdf} only differ by the transition PDF $p_N$. As a result we obtain
	\begin{equation}\label{eq:n-1incpdf}
	\fl p(v_N | v_0,v_1,\dots,v_{N-1},u_0) = \frac{f(v_0,v_1,v_2,v_3,\dots,v_N,u_0)}{f(v_0,v_1,v_2,v_3,\dots,v_{N-1},u_0)} = p_N\big( v_N+u_0 \big| u_0, v_{N-1}+u_0 \big) \, .
	\end{equation}
	This relation shows that only the transition PDF $p_N$ of the $N$th refinement level is needed to fully specify increment PDF conditional on all other increments and $u_0$. In other words, the conditional PDF $p(v_N | v_0,v_1,\dots,v_{N-1},u_0)$ indeed is independent of $v_0,\dots,v_{N-2}$ such that the generalized Markov property\footnote{The Markov property is generalized in the sense that the conditional PDFs maintain a dependence on $u_0$. The necessity of including $u_0$ has also been discussed in a related approach presented in \cite{stresing10njp}.}
	\begin{equation}\label{eq:markovproperty}
	p(v_N | v_0,v_1,\dots,v_{N-1},u_0) = p(v_N | v_{N-1},u_0)
	\end{equation}
	holds. Of course, this argument holds on all larger scales as well, showing that MLSR generates a Markov process in scale.

	\section{Proof-of-Concept: Nonlinear Langevin Equations}\label{sec:data-driven-langevin}
	The method described in section \ref{sec:threepointpdf} can, given Markov properties, produces synthetic time series with correct statistics from any experimental or numerical time series with sufficient statistics to resolve the three-point PDFs $f(u_{\mathrm l},u_{\mathrm c},u_{\mathrm r};r_i)$ at a hierarchy of scales $r_i$. From these three-point PDFs and the corresponding two-point PDFs at each scale, $f(u_{\mathrm l},u_{\mathrm r};2r_i)$, one can then compute the conditional PDFs, $p_i(u_{\mathrm c}|u_{\mathrm l},u_{\mathrm r})$, through equation \eref{eq:conditionalpdf} and use those to refine a coarsely sampled time series at scale $r_0$ down to scale $r_N$. 
	\begin{figure}[h]
		\centering
		\includegraphics[width=1.0\textwidth]{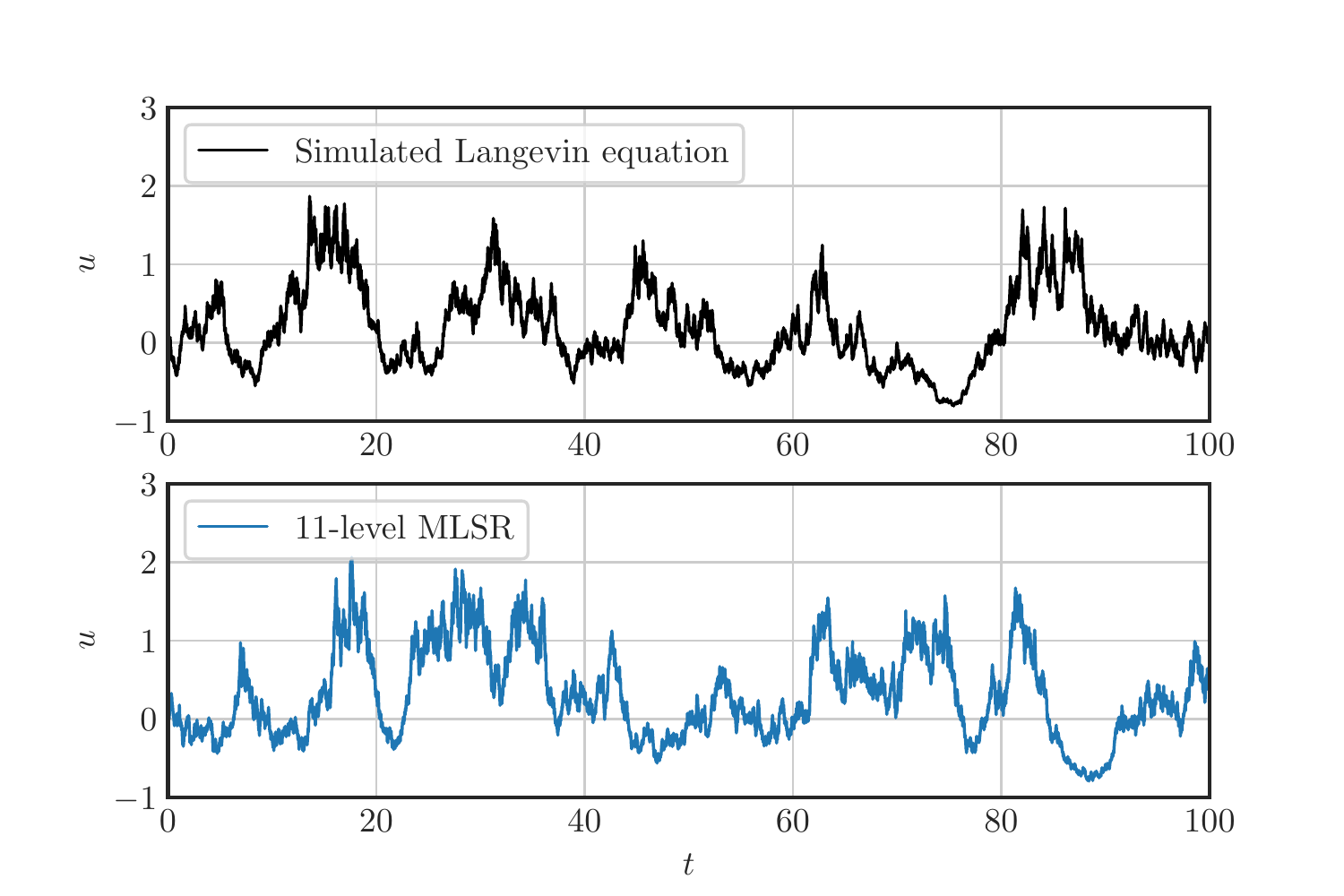}
		\caption{Implementation of the MLSR method to the nonlinear Langevin equation \eref{eq:langevin} with $\gamma = 1$, $c_1=5$, $\sigma = 1$ and $c_2 = 1$. The top plot shows a time series of the simulated system, the bottom plot a 11-level refinement from the MLSR method.}
		\label{fig:langevin_mlsr}
	\end{figure}
The quality of the refinement depends on the statistical properties of the original time series as well as on the convergence of the measured PDFs. In general, the further the original time series deviates from exhibiting Markovian properties, the less accurate the reconstruction becomes.

	As a proof-of-concept test case, let us consider a prototypical nonlinear Langevin equation of the form
	\begin{equation}
	\mathrm{d}u = -\gamma u^{c_1}\mathrm{d}t+\sqrt{\left(\sigma+c_2 u\right)^2}\mathrm{d}W_t.
	\label{eq:langevin}
	\end{equation}

	Here, $\mathrm{d}W_{t}$ is the increment of a Wiener process in time, $\gamma$ is a parameter controlling the strength of the (nonlinear) drift term, and $\sigma$ a diffusion parameter. The parameters $c_1$ and $c_2$ determine the degree of nonlinearities of the respective terms and give rise to non-Gaussian behavior. For $c_1=1$ and $c_2 = 0$ this equation reduces to the standard Ornstein-Uhlenbeck process with linear drift and constant diffusion that results in Gaussian solutions. The system can be directly simulated using a standard Euler-Maruyama method \cite{Kloeden1992}. To obtain sufficient data for the three-point PDFs, we compute a total of $1.2\cdot10^{10}$ data points with a time step of $\Delta t = 10^{-3}$ for various choices of the parameters. We then apply the MLSR method to refine initial conditions at scale $\tau_0 = 2^{11}\Delta t$ down to scale $\tau_{11} = \tau_0/2^{11}$.
	Such a refinement is shown in figure
	\ref{fig:langevin_mlsr}.

	\begin{figure}[h!]
		\centering
		\includegraphics[]{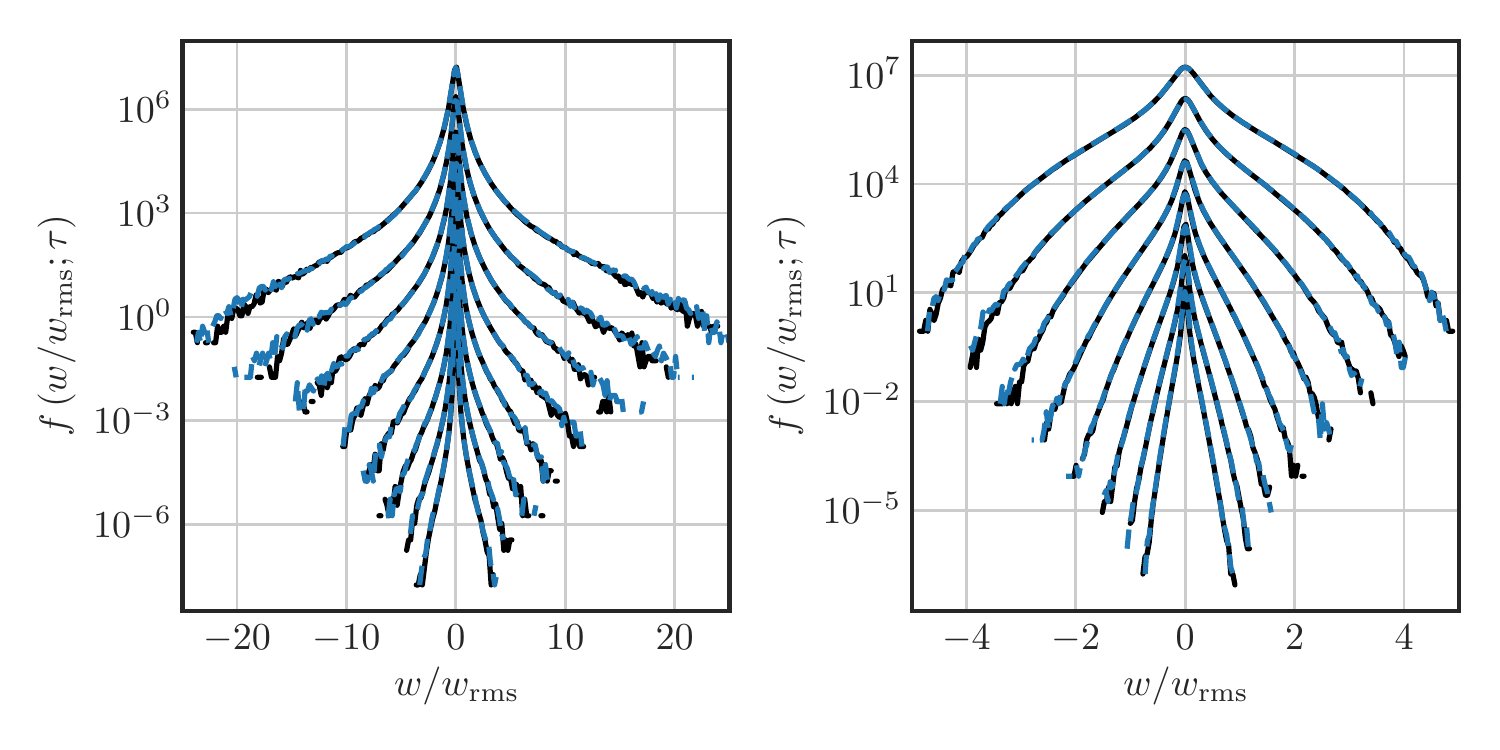}
		\caption{Increment PDFs of simulated (black solid lines) Langevin equation \eref{eq:langevin} and the respective MLSR reproductions (blue dashed lines). The curves are staggered for better visibility corresponding to scales $\tau= \tau_0$ (bottommost) through $\tau = \tau_7$ (topmost). (Left): The parameters are $\gamma = 1$, $c_1=1$, $\sigma = 1$ and $c_2 = 1$. This choice results in a heavy-tailed increment distribution at all scales that is correctly reproduced by the MLSR algorithm. (Right): The parameters are $\gamma = 1$, $c_1=5$, $\sigma = 1$ and $c_2 = 1$. The sub-Gaussian cores of the increment PDFs produced by the strongly nonlinear drift as well as the heavy tails are well-reproduced.}
		\label{fig:inclangevin1111}
	\end{figure}
	The properties of the MLSR reproduction are virtually indistinguishable from the original, simulated time series. This can also be tested quantitatively by the means of the scale-dependent PDFs $f\left(w;\tau\right)$ of the increment $w = u\left(t+\tau\right)-u\left(t\right)$ at a given scale $\tau$. This is shown in figures \ref{fig:inclangevin1111} for two sets of parameters.

	\vspace{1cm}

	\section*{References}

	\bibliographystyle{iopart-num.bst}
	\bibliography{bib.bib}

\end{document}